\pdfoutput=1
\documentclass[11pt]{article}

\usepackage{times}
\usepackage{latexsym}
\usepackage[T1]{fontenc}

\usepackage[utf8]{inputenc}

\usepackage{microtype}

\usepackage{inconsolata}

\usepackage{graphicx}

\usepackage[inline,shortlabels]{enumitem}
\usepackage{tabularx}
\usepackage{bigstrut}
\usepackage{rotating}
\usepackage{amsmath}
\usepackage{algorithm}
\usepackage{algpseudocode}
\usepackage{booktabs}
\usepackage{multirow}

\PassOptionsToPackage{bookmarks=false,colorlinks=false,hidelinks}{hyperref}

\usepackage[final]{acl}

\title{QUIDS: 
Query Intent Description for Exploratory Search\\ via Dual Space Modeling}

\author{
  Yumeng Wang \\
  Leiden University \\
  Leiden, The Netherlands \\
  {\small \texttt{y.wang@liacs.leidenuniv.nl}}
  \And
  Xiuying Chen \\
  MBZUAI \\
  Abu Dhabi, United Arab Emirates \\
  {\small \texttt{xiuying.chen@mbzuai.ac.ae}}
  \And
  Suzan Verberne \\
  Leiden University \\
  Leiden, The Netherlands \\
  {\small \texttt{s.verberne@liacs.leidenuniv.nl}}
}

\begin{document}
\maketitle
\begin{abstract}
In exploratory search, users often submit vague queries to investigate unfamiliar topics, but receive limited feedback about how the search engine understood their input. This leads to a self-reinforcing cycle of mismatched results and trial-and-error reformulation. 
To address this, we study the task of generating user-facing natural language query intent descriptions that surface what the system likely inferred the query to mean, based on post-retrieval evidence.
We propose QUIDS, a method that leverages dual-space contrastive learning to isolate intent-relevant information while suppressing irrelevant content. QUIDS combines a dual-encoder representation space with a disentangling decoder that works together to produce concise and accurate intent descriptions.
Enhanced by intent-driven hard negative sampling, the model significantly outperforms state-of-the-art baselines across ROUGE, BERTScore, and human/LLM evaluations. Our qualitative analysis confirms QUIDS' effectiveness in generating accurate intent descriptions for exploratory search.
Our work contributes to improving the interaction between users and search engines by providing feedback to the user in exploratory search settings.\footnote{Our code is available at \url{https://github.com/menauwy/QUIDS}}

\end{abstract}

\section{Introduction}

\begin{figure}[t]
\includegraphics[width=\columnwidth]{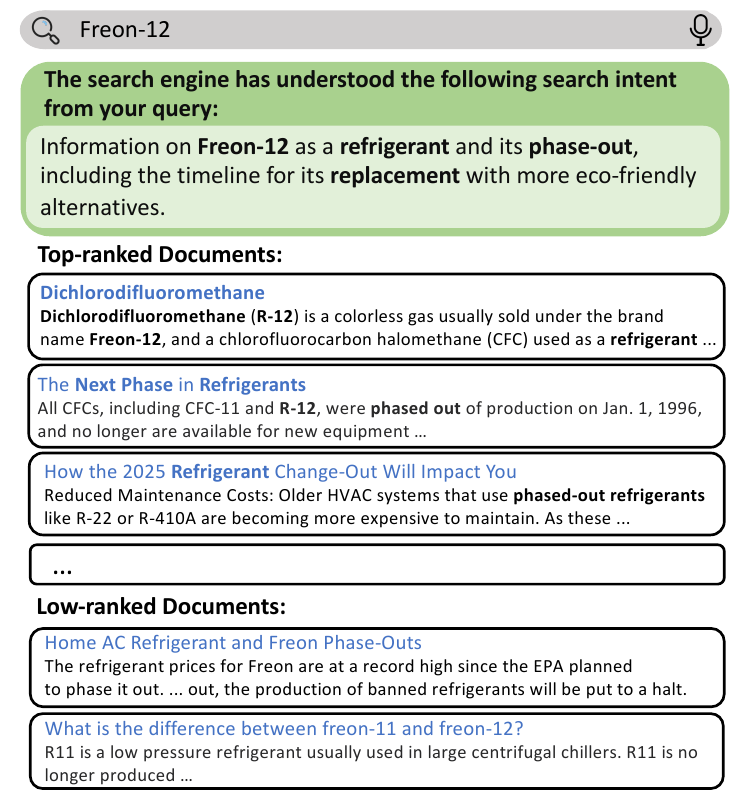}
\caption{A user-facing application of query intent generation in exploratory search. The system's inferred intent is generated by contrasting top-ranked (pseudo-relevant) and low-ranked (pseudo-irrelevant) documents. Key information contributing to the inferred intent is shown in bold.}
\label{fig:figure1}
\end{figure}

In exploratory search~\citep{palagi2017survey}, users often issue vague or underspecified queries to investigate unfamiliar topics through iterative refinement. This process gives rise to a persistent usability challenge, which we call the \textit{dual-blind problem}: Users are uncertain about how to express their information needs; as a result they formulate ambiguous queries; the system silently infers the user's intent based on these ambiguous queries, without providing explicit feedback and retrieves mixed-quality results at best. This leads to a self-reinforcing cycle of the user receiving results that are misaligned with their actual information need. 
This cycle is difficult to break with traditional query understanding methods~\cite{li2023graph, lu2024effective, wang2023query2doc}
that operate in the \textit{pre-retrieval} stage. Their goal is to optimize ranking effectiveness, not to provide feedback to the user. 
They offer little transparency about how the system arrived at its results, which is especially a problem  when users are unsure of their intent.

In this work, we study the task of generating a natural-language query intent description~\citep{zhang2020q2id} that reflects what the system likely inferred the query to mean. The description is generated in the \textit{post-retrieval} stage, incorporating system-inferred relevance of documents. 
These descriptions are not mechanistic explanations of the ranker, but instead serve as user-facing proxies of the system’s inferred intent. By contrasting top-ranked (pseudo-relevant) and low-ranked documents, the intent descriptions 
provide feedback that helps users identify mismatches between their intended and inferred query meanings. This feedback supports more effective query refinement and improves the overall search experience. 

Figure \ref{fig:figure1} illustrates an application scenario of query intent generation in exploratory search. Since the ground truth intent behind the query is unknown to the search system, it relies on the retrieved documents to infer the query intent. 
From the top-ranked documents (considered relevant by the search engine), key terms like ‘Freon-12’, ‘refrigerant’, and ‘phased out’ are captured and emphasized in the intent description. In contrast, topics such as ‘costs’ and ‘prices’, which appear in both high-ranked and low-ranked documents, are excluded from the final intent description.
The resulting description provides diagnostic feedback on how the system understood the query, helping users assess whether the retrieved results align with their latent intent.

We introduced a novel dual-space modeling approach, QUIDS, for the query intent generation task. 
It models query intent through dual-space contrastive learning by performing contrastive learning in two complementary spaces, explicitly separating intent-relevant and irrelevant semantic information. Specifically, the method consists of:
(i) a \textit{representation space} via dual encoders and (ii) a novel \textit{disentangling space} in the decoder. This dual-space design enables the model to subtract irrelevant semantics from relevant ones, generating concise and more accurate intent descriptions. Furthermore, we propose an intent-driven hard negative sampling strategy to expand the irrelevant representation space and improve contrastive learning during training.

Experiments on the Q2ID benchmark~\citep{zhang2020q2id}, including TREC and SemEval datasets, show that our model significantly outperforms strong baselines, including the prior Q2ID-specific method, LLM-based, and Query-focused Summarization methods, both in automatic and human evaluations. Qualitative analysis confirms that QUIDS effectively filters out distracting or misleading content and generates concise intent descriptions. Our contributions are: 
\begin{enumerate*}[label=(\roman*)]
    \item Our model generates high-quality intent descriptions, with performance significantly enhanced by incorporating hard negative data augmentation during training.
    \item We introduce contrastive learning in both the representation space and the disentangling space of transformer models, effectively capturing contrasting information from relevant and irrelevant documents.
    \item We perform a thorough evaluation of our model, 
    providing us with insights into the model’s strengths and weaknesses, as well as its potential application scenarios, especially for exploratory search. 
\end{enumerate*}






\section{Related Work}

\subsection{Query Understanding}
Our work is related to traditional query understanding tasks such as classification~\cite{broder2002taxonomy,verberne2013reliability}, clustering~\cite{wen2002query, hong2016accurate}, and expansion~\cite{wang2023query2doc, mo2023convgqr, jagerman2023query}. 
However, unlike these methods, which operate in the \textit{pre-retrieval} stage to optimize retrieval effectiveness, \citet{zhang2020q2id} proposes the Query-to-Intent-Description (Q2ID) task in the \textit{post-retrieval} stage that aims to generate search systems' inferred intent of a user query based on both relevant and irrelevant documents.
Unlike their method, we directly model a query-aware irrelevant intent space via dual-space contrastive learning, and enhance the performance with hard negative data augmentation, leading to a more precise intent description.

\subsection{Query-focused Summarization}
In settings where annotated intent descriptions are available, a related task to Q2ID is query-focused summarization (QFS)~\citep{vig2022exploring,pagnoni-etal-2023-socratic}. 
QFS is a subtask of text summarization that aims to generate a summary of one or multiple documents, guided by a query.
Traditional methods rely on unsupervised extraction, ranking text segments by similarity and query relevance \citep{wan2009graph, feigenblat2017unsupervised}. Recent QFS datasets \citep{kulkarni2020aquamuse, fabbri-etal-2022-answersumm, zhong2021qmsum} have enabled the rise of QA-driven approaches \citep{su-etal-2020-caire, su2021improve}.
More advanced techniques model query relevance through evidence ranking \citep{xu2021generating}, latent query optimization \citep{xu2022document}, or pipeline architectures like the coarse-to-fine model in \cite{xu2020query}. 
To handle long documents, extract-then-abstract strategies \citep{vig2022exploring} use sparse attention and segment scoring. Other innovations include question-driven pretraining \citep{pagnoni-etal-2023-socratic}, contrastive learning \citep{sotudeh2023qontsum}, and joint token-utterance modeling with query-aware attention \citep{liu2023query}.

We use QFS models as baselines for the Q2ID task, but there is a fundamental difference between QFS and Q2ID: QFS aims to compress the content of retrieved documents to help users consume information, whereas Q2ID aims to generate a description of what the system likely inferred about the query intent, based on retrieval results. 
We provide a comparison table with related tasks in Appendix~\ref{app:related_work}.

\section{Methods}

\begin{figure*}[t]
    \centering
    \includegraphics[width=0.9\textwidth]{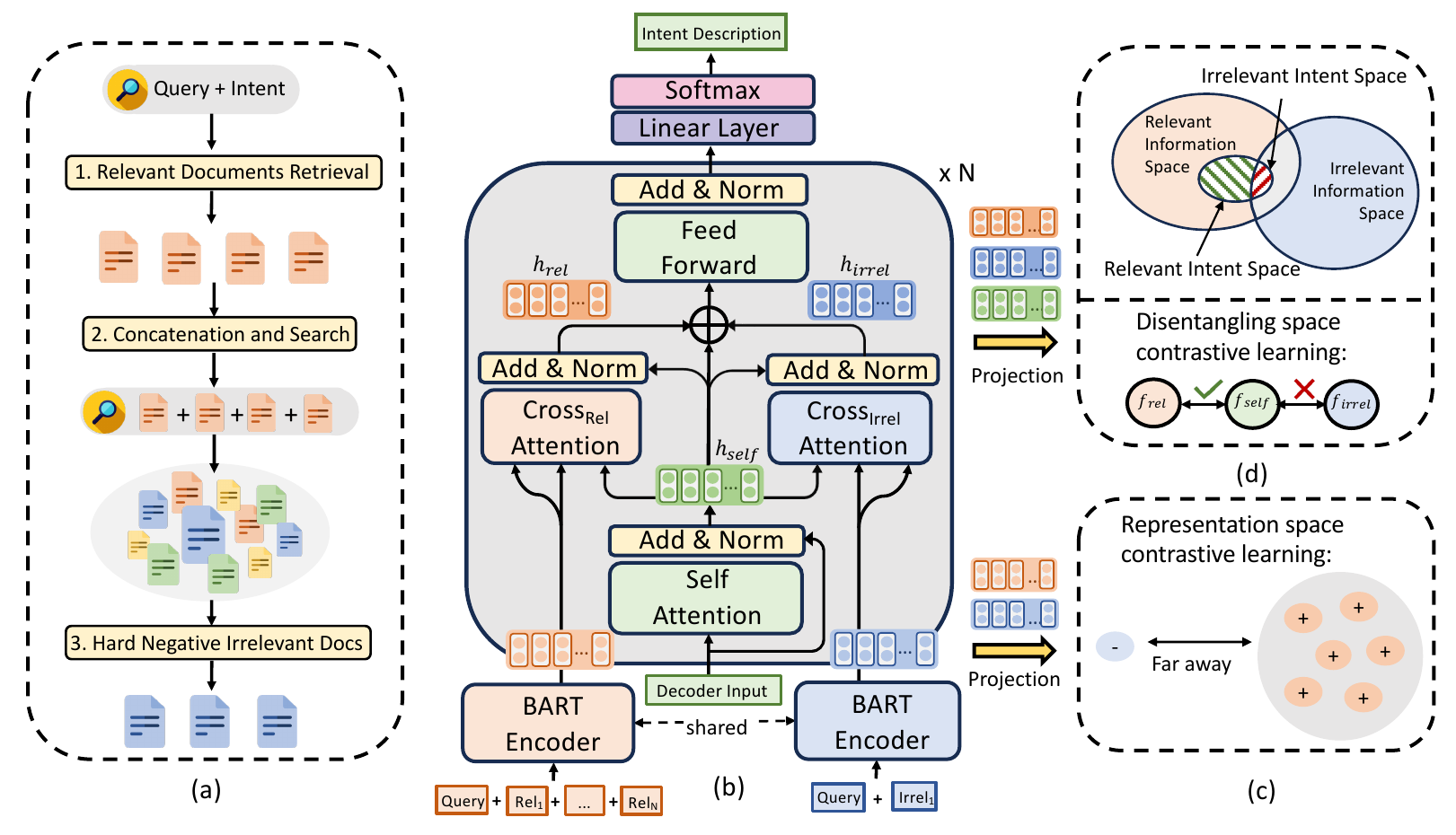}
    \caption{Overview of our proposed pipeline. From left to right, we show (a) Intent-Driven Negative Augmentation method, (b) Contrastive decoder structure with dual cross-attention layers, (c) and (d) Contrastive learning via dual space modeling.}
    \label{fig:pipeline}
\end{figure*}


\subsection{Pipeline Framework}
We define the contrastive intent generation task 
as follows. Given a dataset $\mathcal{D} = \{(q, \mathcal{R}, \mathcal{I}, y)_j\}$ with $L$ samples, where $j \in \{0, 1, ..., L\}$:  
$q$ is a query, $\mathcal{R} = \{r_1, r_2, ..., r_{|\mathcal{R}|}\}$ is a collection of relevant documents for the query, $\mathcal{I} = \{i_1, i_2, ..., i_{|\mathcal{I}|}\}$ is a collection of irrelevant documents and $y$ is the human-annotated ground truth query intent.
The modeling goal is to learn the distinctions between relevant and irrelevant inputs based on a query, while generating a system-inferred intent description that exclusively highlights the relevant aspects related to the query.
To achieve this, our training pipeline consists of 2 steps: (1) \textit{Intent-Driven Negative Augmentation (IDNA)} and  
(2) \textit{Dual Space Modeling (DualSM)}.

\subsection{Intent-Driven Negative Augmentation (IDNA)}\label{IDNA}

The purpose of IDNA is to mine hard negative documents as irrelevant documents from the entire dataset $D$ based on the query, its relevant document collections, and the ground truth intent, i.e.,
\[IDNA(q, \mathcal{R}, y, \mathcal{D}) = \mathcal{I}'\]
where $\mathcal{I}' = \{i'_1, i'_2, ..., i'_h\}$ with $h$ the expected number of irrelevant documents.
Inspired by \citet{liu-etal-2022-makes} on choosing in-context sample strategies for in-context learning, we design a method to choose intent-aware hard negative samples based on semantic similarity.
Specifically, we use a Sentence Transformer model \citep{reimers-2019-sentence-bert} to represent both positive and negative samples from the training data in a vector space. Then we choose negative samples close to the positive ones in this space. The negative sample augmentation makes the task more challenging for the model, hence improving its discriminative capabilities. 
As shown in Figure~\ref{fig:pipeline} (a), 
we augment hard negative samples for each training query using Algorithm~\ref{alg:IDNA}.
\captionsetup[algorithm]{font=small}
\begin{algorithm}[t]
\small
\caption{Intent-Driven Negative Augmentation (IDNA)}\label{alg:IDNA}
\begin{algorithmic}[1]
\Require query $q$,  relevant document collection $\mathcal{R}$, irrelevant document collection $\mathcal{I}$, whole dataset document corpus $\mathcal{D}$, target size $S$, threshold $\tau$
\Ensure augmented irrelevant documents $\mathcal{I}'$ for $q$
\State $h_{q;y} \gets \text{Encode}(\text{Concatenate}(q; y))$
\State $\mathcal{R}^* \gets \text{Sort } \mathcal{R} \text{ descending by } \cos(h_{q;y}, \text{Encode}(r)),$
\Statex $\forall r \in \mathcal{R}$
\State $h_{R^*} \gets \text{Encode}(\text{Concatenate}(\mathcal{R}^*))$ 
\State $\mathcal{I}^* \gets \text{Sort } \mathcal{I} \text{ descending by } \cos(h_{R^*}, \text{Encode}(I)),$
\Statex $\forall i \in \mathcal{I}$ 
\State $\mathcal{I}' \gets$ top-$|\mathcal{I^*}|$ ranked docs
\While{$|\mathcal{I}'| < S$} 
    \State Sample a document $d$ from $\mathcal{D}$
    \If{$\cos(h_{R^*}, \text{Encode}(d)) > \tau$}
    \State Add $d$ to $\mathcal{I}'$
    \EndIf
\EndWhile
\end{algorithmic}
\end{algorithm}
For encoding, we use a Sentence Transformers model pre-trained on the MSMARCO Passage Ranking dataset \citep{nguyen2016ms}.
In practice, we set 0.8 to the similarity threshold with an analysis of its effect in subsection~\ref{subsub:IDNA}.

\subsection{Dual Space Modeling (DualSM)}
Dual Space Modeling aims to contrastively generate a descriptive intent for the query by modeling query-aware relevant and irrelevant intent spaces:
\[DualSM(q, \mathcal{R}, i') = \hat{y}\]

We use a Transformer-based encoder-decoder architecture, with the BART-large model \citep{lewis-etal-2020-bart} and the T5-large  model \cite{raffel2020t5model} as backbones, as illustrated in Figure~\ref{fig:pipeline}(b). 
To capture the relationship between a query and its relevant and irrelevant documents, we implement a Siamese dual encoder architecture. Based on the encoder outputs, contrastive learning is performed in the representation space to differentiate between embeddings for relevant and irrelevant documents.
Correspondingly, we design a contrastive decoder to model query-aware relevant and irrelevant intent spaces in a disentangled manner. 


\subsubsection{Representation Space Modeling (RSM)}
We design a dual cross-encoder architecture to distinguish relevant and irrelevant documents for a given query by jointly encoding each query-document pair. Relevant documents, which often share similar topics tied to the query’s intent, are concatenated based on their ranking (Section~\ref{IDNA}, Step 1) and encoded together. In contrast, irrelevant documents can be irrelevant to a query in diverse ways, making it impractical to model a meaningful and comprehensive irrelevant feature representation space. 
Therefore, we focus on a single irrelevant document $i'$ at each training step, using a hard negative sample from the augmented irrelevant document collection $\mathcal{I'}$. 

To model the feature space, we project the encoder’s final hidden states through a linear layer. Document embeddings are obtained via average pooling over token representations. We optimize the representation space by pulling relevant embeddings closer and pushing the irrelevant one away (Figure~\ref{fig:pipeline}(c)). The objective is to minimize:
\begin{equation}
\mathcal{L}_{rel} =\textstyle  \sum^{k}_{m=1} \sum^{k}_{n=m+1} d(e_m, e_n)
\end{equation}
where $e$ is the embedding of each relevant document, $k$ is the number of relevant documents, and $d$ is a distance function. We use cosine distance for $d$ in this work. 
For irrelevant feature representation space, we optimize the margin loss function:
\begin{equation}
\mathcal{L}_{irrel} = \textstyle \sum^{k}_{m=1} max(t-d(e_m, \bar{e}),0)
\end{equation}
where $\bar{e}$ is the embedding of the irrelevant document, and $t$ is a margin parameter, set to 1 in our case. 
We combine the relevant and irrelevant loss to obtain the encoder contrastive loss as follows:
\begin{equation}
\mathcal{L}_{encoder} = \mathcal{L}_{rel} + \mathcal{L}_{irrel}
\end{equation}

\subsubsection{Disentangling Space Modeling (DSM)}
In decoding, we aim to generate intent descriptions based on the encoded relevant and irrelevant document features. To achieve this, we design a contrastive decoder with an added cross-attention layer that attends to both sources. To further disentangle relevant from irrelevant information, we apply contrastive learning in a separate disentangling space. As shown in Figure~\ref{fig:pipeline}(d), this helps the model focus on relevant intent while minimizing influence from irrelevant content, enabling more precise and nuanced intent generation.

Our decoder adopts a Transformer architecture, composed of N identical decoder layers. In the $l$-th decoder layer, at the $z$-th decoding step, we obtain hidden states $h_{self,z}^{l}$ by employing masked self-attention layers, to make sure the prediction of position $z$ depends only on the predictions before $z$.
Based on $h_{self, z}^{l}$, we compute relevant document hidden states $h_{rel,z}^{l}$ by applying multi-head attention with cross-attention (MHAtt) to relevant encoder output:
\begin{equation}
h_{rel,z}^{l} = MHAtt(h_{self,z}^{l}, h_{R^*})
\end{equation}
Similarly, we get the irrelevant document hidden states by attending to irrelevant encoder output:
\begin{equation}
h_{irrel,z}^{l} = MHAtt(h_{self,z}^{l}, h_{i'})
\end{equation}
From preliminary results, we found that a simple linear combination of $h_{self,z}^{l}$, $h_{rel,z}^{l}$, and $h_{irrel,z}^{l}$ works well to serve as the decoder hidden state to produce the distribution over the target vocabulary:
\begin{equation}
h_{combine,z}^{l} = h_{self,z}^{l}+h_{rel,z}^{l}-h_{irrel,z}^{l}
\end{equation}
\begin{equation}
P^{vocab}_z = Softmax(W(h_{combine,z}^{N}))
\end{equation}
where $W$ indicates a linear transformation.
We optimize the model with the negative log likelihood (NLL) objective to predict the target words:
\begin{equation}
\mathcal{L}_{NLL} = - \textstyle \sum^{|y|}_{z=1} log P^{vocab}_z(y_z)
\end{equation}
Corresponding to the representation space contrastive learning, we perform another contrastive learning in the newly proposed disentangling space using hidden states from the last decoder layer. 
We apply an additional linear layer to $h_{self}^{N}$, $h_{rel}^{N}$, and $h_{irrel}^{N}$, projecting them into a new representation space. We then obtain the embeddings $f_c$, $f_r$, $f_{i'}$ by pooling these projected vector representations.

We follow the approach of SimCLR \citep{chen2020simple} and use from-batch negative samples $\mathcal{B}$ in the InfoNCE loss \citep{he2020momentum}:
\begin{equation}
\mathcal{L}_{decoder} = -log \frac{exp(cos(f_{c},f_{r})/\tau)}{\sum_{i' \in \mathcal{B}} exp(cos(f_{c}, f_{i'})/\tau)}
\end{equation}
where $\tau$ is the temperature and $cos(\cdot, \cdot)$ defines cosine similarity.

Finally, we combine the original NLL loss together with encoder and decoder loss to obtain the overall loss $\mathcal{L}$ to update all learnable parameters in an end-to-end learning setting:
\begin{equation}
\mathcal{L}_{NLL} = \lambda_0\mathcal{L}_{NLL} + \lambda_1\mathcal{L}_{Encoder} + \lambda_2\mathcal{L}_{Decoder}
\end{equation}
where the $\lambda$ parameters control the balance between the three losses, with their total sum equal to 1.

\section{Experimental Settings}


\subsection{Data}
We conduct experiments on the \textbf{Q2ID} dataset \citep{zhang2020q2id}, a benchmark for query-to-intent description derived from existing TREC and SemEval collections. Specifically, it comprises:
\textbf{TREC}: Including the Dynamic Domain tracks (2015–2017) and the 2004 Robust Track, which focus on dynamic, exploratory search and consistency of retrieval technology.
\textbf{SemEval}: Including the English SemEval-2015 and SemEval-2016 Task 3 tracks on Community Question Answering.
Q2ID contains a total of 5,358 entries. Each entry is structured as a quadruple: \textit{<query, relevant documents, irrelevant documents, intent description>}, where the intent descriptions are human-written narratives.
The statistics and more details are provided in Appendix~\ref{appendix:dataset-details}.

\subsection{Baselines}\label{baselines}
To reflect the shared focus on user queries and the extraction of relevant content, we compare our model with baselines from four categories: (i)~\textit{Pretrained Seq2Seq Models}: We fine-tune \textbf{T5-large} \citep{raffel2020t5} and \textbf{BART-large} \citep{lewis-etal-2020-bart} on the Q2ID dataset. BART also serves as the backbone of our QUIDS model.
(ii)~\textit{Q2ID Baseline}: \textbf{CtrsGen} \citep{zhang2020q2id} leverages contrastive generation using a bi-GRU encoder and contrast-weighted attention mechanism.
(iii)~\textit{LLM Baseline}: We evaluate a instruction-tuned model \textbf{LLaMA3.1-8B-Instruct} \citep{llama31} and a reasoning model \textbf{OpenAI o3} \citep{openai2025o3} in zero-shot and two-shot settings. For the two-shot settings, examples are drawn from TREC and SemEval.
(iv)~\textit{QFS Baselines}: We include extractive-abstractive models \textbf{RelReg}, \textbf{RelRegTT}~\citep{vig2022exploring}, the segment-based model \textbf{SegEnc}~\citep{vig2022exploring}, the question-driven \textbf{Socratic} \citep{pagnoni-etal-2023-socratic}, and the contrast-enhanced \textbf{Qontsum} \citep{sotudeh2023qontsum}.
Detailed descriptions, training configurations, and reproduction settings for baselines are in Appendix~\ref{app:baseline_details} and \ref{app:implementation_baselines}.

\subsection{Implementation Details}
Our method is implemented based on the BART-large model \citep{lewis-etal-2020-bart} using the Huggingface Transformers library \citep{wolf2019huggingface}. Cross\textsubscript{Rel} and Cross\textsubscript{Irrel} attention layers in the decoder are initialized with pre-trained BART weights.
We optimize the weighted training loss using coefficients $(\lambda_0=0.2, \lambda_1=0.2, \lambda_2=0.6)$ to balance multiple objectives (see Appendix~\ref{appendix:loss-weights}). The model is trained with the Adam optimizer, and the final checkpoint is selected based on average ROUGE-\{1, 2, L\} scores on the validation set.
We provide additional training details in 
Appendix~\ref{appendix:impl-details}.

\subsection{Evaluation Metrics}
We conduct three types of evaluations using different evaluator resources: automatic evaluation, LLM-based evaluation and human evaluation. 
For automatic evaluation, we report recall scores on ROUGE-\{1, 2, L\} following \citet{zhang2020q2id}, along with BERTScore \citep{zhang2019bertscore}, which assesses semantic and syntactic similarity beyond exact word matches.
We also conduct a human evaluation study using 50~\citep{sotudeh2023qontsum} randomly selected test samples. 
Five PhD students in Computer Science scored intent descriptions from our model and the best baseline, without knowing which model produced them. They rated both models on four customized qualitative criteria with scores ranging from 1 (worst) to 5 (best). 
Four criteria are:
(1) \textbf{Fluency}: to what extent the generated query intent description reads naturally, understandably, and without noticeable errors or disruptions.
(2) \textbf{Factual Alignment}: to what extent the generated query intent description is factually aligned with the ground truth intent.
(3) \textbf{Inclusion score}: how well the generated query intent includes important details from the query and relevant documents.
(4) \textbf{Exclusion score}: how well the generated query intent description excludes information present in the irrelevant documents that is not relevant to the query and relevant documents.
Inspired by \citet{liu-etal-2023-g}, we also adopt LLM-based evaluation (LLaMa3.1-8B-Instruct and GPT-4o) by prompting instruction-tuned models to assess generations across the same four qualitative metrics.
Details on the prompt formats, and scoring computation are provided in Appendix~\ref{appendix_prompt}.

\begin{table}[t]
    \centering
    \resizebox{\linewidth}{!}
    {
    \begin{tabular}{lcccc}
    \toprule[1pt]  
    \textbf{Models} & \textbf{RG-1} & \textbf{RG-2} & \textbf{RG-L} &\textbf{BS} \\
    \midrule
    CtrsGen\textsuperscript{$\dagger$} & 24.76 & 4.62 & 20.21 & - \\
    T5-large & 28.87 & 13.91& 23.85 & 61.64 \\
    BART-large & 30.70 & 13.91 & 24.63 & 62.07 \\
    \midrule
    LLaMa3.1 (0) & 29.28 & 7.42& 20.90 & 57.26\\
    LLaMa3.1 (2) & 32.75 & 9.54 & 24.34 & 57.89\\
    o3 (0) & \textbf{34.91} & 6.84 & 23.93 & 59.19\\
    o3 (2) & 33.15 & 6.28 & 22.81 & 59.88\\
    \midrule
    RelReg & 26.67& 12.83 & 21.99 & 59.24\\
    RelRegTT & 27.21 & 12.77 & 22.25 &  59.60\\
    SegEnc & 31.83 & \underline{14.29} & \underline{25.18} & 62.15 \\
    \small + SOCRATIC Pret. & 31.38& 13.88 & 24.91& \underline{62.26}\\
    QONTSUM & 31.18 & 14.26 & 24.87 & 62.03 \\
    \midrule
    QUIDS\_T5 & 29.40 & 13.95 & 24.23 & 62.00 \\
    QUIDS\_BART & \underline{34.47}& \textbf{14.86\textsuperscript{$*$}} & \textbf{26.77\textsuperscript{$*$}}& \textbf{63.55\textsuperscript{$*$}}\\ 
    \bottomrule[1pt]
    \end{tabular}
    }
    \caption{Performance between our model and baselines in terms of automatic evaluation (\%). \textsuperscript{$\dagger$} indicates reported performance from previous work. `-' means the result is inaccessible. \textsuperscript{$*$} indicates the model outperforms the best baseline significantly with paired t-test at $p$-value $<$ 0.05 level. Results are averaged over 5 random seeds. The best results are highlighted in bold, while the best baseline results are underlined.}
    \label{tab:benchmark}
\end{table}
\begin{table}[t]
\centering
\resizebox{\linewidth}{!}
{
\begin{tabular}{lcccc}
\toprule[1pt]
\textbf{Model} & \textbf{RG-1} & \textbf{RG-2} & \textbf{RG-L} &\textbf{BS} \\
\midrule
QUIDS w/o IDNA & 33.48 & 14.20 & 25.95 & 63.17 \\
QUIDS w/o RSM & 34.57 & 14.39 &  26.38 &  63.62 \\
QUIDS w/o DSM & 33.45 & 13.46 & 25.88 & 63.33 \\
\midrule
QUIDS & \textbf{35.95}& \textbf{14.80} & \textbf{27.21}& \textbf{64.33}\\ 
\bottomrule[1pt]
\end{tabular}
}
\caption{Ablation study of our QUIDS model with its variants under automatic evaluation (\%).}
\label{tab:ablation_bart-large}
\end{table}

\begin{figure*}[t]
    \centering
    \includegraphics[width=\textwidth]{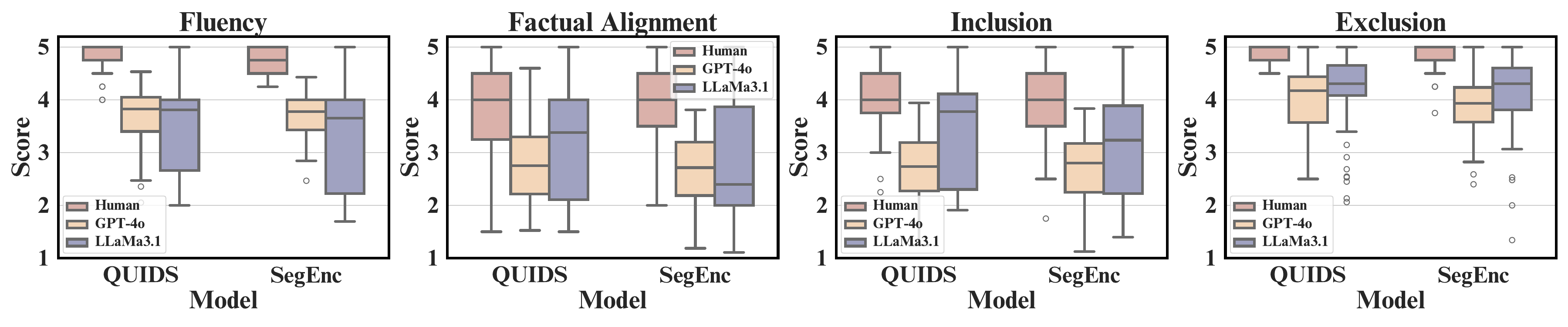}
    \caption{Distribution of human and LLM evaluation sores on four qualitative metrics.}
    \label{fig:plot_evaluation_all}
\end{figure*}

\section{Experimental Results}

\subsection{Overall Results}
We compare model performance between QUIDS and baselines in Table~\ref{tab:benchmark}.
The results show that: (1) QUIDS outperforms all baselines except o3 on RG-1, including those with larger model sizes, indicating that the gains stem primarily from our dual-space modeling design rather than model size or pretraining knowledge; (2) Our approach is compatible with both T5 and BART architectures.
Notably, BART-large outperforms T5-large despite having nearly half the model size;
(3) The QFS models that we implemented for the query intent generation task outperform the Q2ID-specific baseline, CtrsGen. QUIDS further significantly 
outperforms the best QFS model, SegEnc;
(4) The two-shot setting with the LLaMa3.1-8B-Instruct model significantly outperforms the zero-shot setting in ROUGE scores, while showing only minor improvements in BERTScore. This suggests that without fine-tuning, generated intents may be lexically similar but semantically misaligned;
(5) While o3 shows strong reasoning and generates richer descriptions, it does not outperform our model on ROUGE and BERTScore. We observed that o3 often paraphrases intent with different wording, leading to lower ROUGE scores. Its longer outputs and inclusion of detailed justifications or summaries may dilute the concise intent signal. This suggests that strong reasoning alone may not align well with the goal of generating concise, system-inferred intent.

\subsection{Ablation Study}
We perform an ablation study based on the BART-Large model to evaluate the contribution of key components in our approach under three settings: (1) without the IDNA module (w/o IDNA), (2) without contrastive learning in the encoder (w/o RSM), and (3) without contrastive learning in the decoder (w/o DSM). 
Results are shown in Table~\ref{tab:ablation_bart-large}.
Excluding contrastive learning from the decoder leads to the largest performance drop, underscoring its role in modeling a discriminative intent space. Removing it from the encoder results in a smaller decline, suggesting that representation space modeling still contributes to relevance awareness.
\begin{figure*}[t]
    \includegraphics[width=\textwidth]{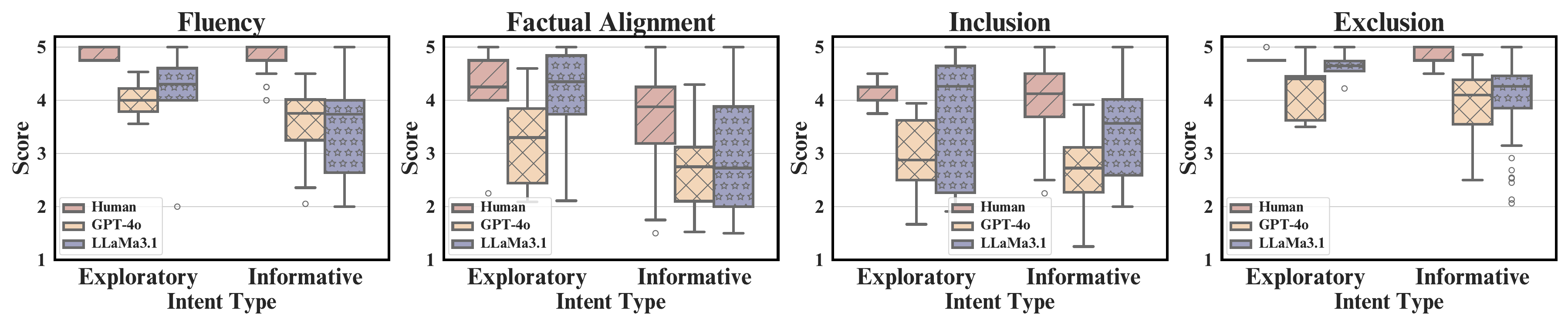}
    \caption{Boxplot of human and LLM evaluation scores on 4 metrics of our model on different intent types.}
    \label{fig:plot_intent_type}
\end{figure*}

\subsection{Human and LLM Evaluation}\label{sec:llmeval}
We assess the quality of generated intents from QUIDS and the best baseline SegEnc using both human and LLM-based evaluations. Inter-annotator agreement, measured by weighted Cohen's $\kappa$, and LLM-human correlations, measured by Spearman and Kendall $\tau$, indicate fair to moderate consistency across metrics (Table~\ref{tab:correlation}, Appendix~\ref{app:correlation}). 
As shown in Table~\ref{tab:evaluation}, QUIDS outperforms SegEnc on all metrics except Factual Alignment, where humans prefer SegEnc. 
Further analysis (subsection~\ref{subsec:intent_type}) shows this stems from the dominance of informational queries, on which SegEnc performs better. QUIDS, by contrast, performs better on exploratory queries.
Figure~\ref{fig:plot_evaluation_all} further illustrates score distributions, revealing three key insights:
\begin{table}[t]
\centering
\resizebox{\linewidth}{!}
{
\begin{tabular}{l|l|cccc}
\toprule[1pt]
\textbf{Method} & \textbf{Model} & \textbf{Fluen.} & \textbf{Align.} & \textbf{Inclu.} & \textbf{Exclu.} \\
\midrule[1pt]
\multirow{2}{4em}{Human} & SegEnc & 4.75 & \textbf{3.90}& 3.94 & 4.77\\
 & QUIDS & \textbf{4.80} & 3.80 & \textbf{4.06} & \textbf{4.80}\\
 \midrule
\multirow{2}{4em}{GPT-4o} & SegEnc & 3.70 & 2.64& 2.67& 3.83\\
& QUIDS &  \textbf{3.71} & \textbf{2.79} &  \textbf{2.69} & \textbf{4.00}\\
\midrule
\multirow{2}{4em}{LLaMa3.1} & SegEnc & 3.25 & 2.91& 3.17 & 4.07\\
& QUIDS & \textbf{3.48} & \textbf{3.17} & \textbf{3.42} & \textbf{4.11}\\

\bottomrule[1pt]
\end{tabular}
}
\caption{Comparison of human evaluation and LLM evalaution in terms of Fluency, Factual Alignment, Inclusion score and Exclusion score.}
\label{tab:evaluation}
\end{table}
    (1) Human scores are generally higher than LLM scores, especially for Fluency and Exclusion. The larger variability for fluency scores suggests humans may tolerate minor fluency issues. 
    (2) Human evaluations show broader and lower score distributions for Factual Alignment and Inclusion, aligning more with LLaMa3.1 (Table~\ref{tab:correlation}). In contrast, they mirror GPT-4o’s narrower distribution for Fluency and Exclusion, where correlation is higher. This suggests that evaluators differ in how they assess each metric.
    (3) Fluency and Factual Alignment show stronger alignment between LLM and human evaluations, likely due to being less context-dependent. In contrast, Inclusion and Exclusion scores exhibit weaker correlations, indicating inconsistencies in evaluating context-sensitive criteria.

\begin{table}[t]
\centering
\resizebox{0.9\linewidth}{!}{
\begin{tabular}{lcccc}
\toprule[1pt]
\textbf{Intent} & \textbf{RG-1} & \textbf{RG-2} & \textbf{RG-L} &\textbf{BS} \\
\midrule
Informational &  35.69 & 14.51 & 26.82 & 63.88 \\
Exploratory & \textbf{41.55} &  \textbf{23.24}&  \textbf{38.28} & \textbf{76.65} \\
\bottomrule[1pt]
\end{tabular}}
\caption{Comparison of automatic evaluation on our model for different intent types.}
\label{tab:rouge_intent_type}
\end{table}
\begin{figure*}[!t]
    \centering
    \includegraphics[width=\textwidth]{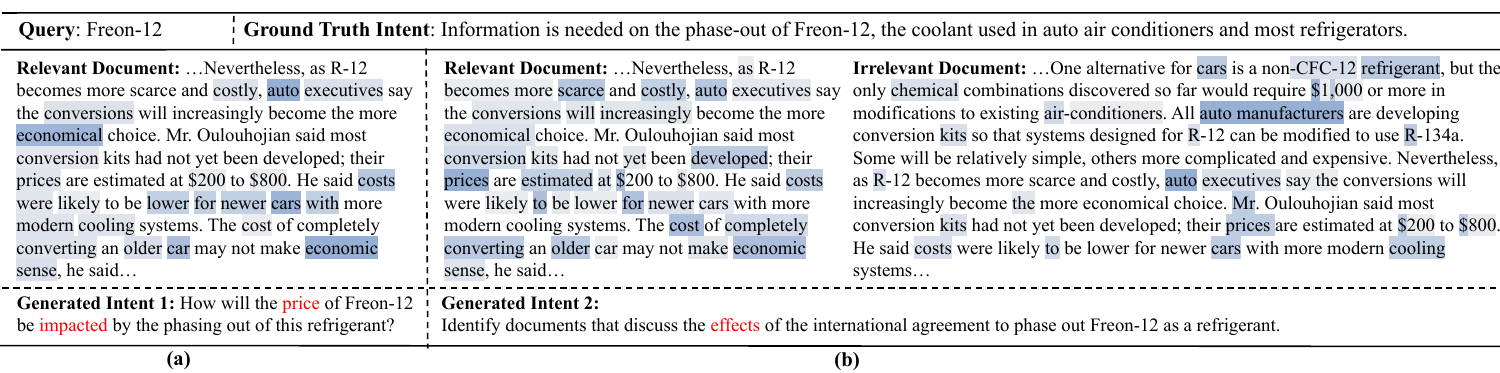}
    \caption{Case study indicating the role of contrastive examples in the decoder stage. Token-level decoder cross-attention weights are shown for a generated intent token (red) are shown with (a) and without (b) an irrelevant document in the model input. Deeper color indicates a higher value.
    }
    \label{fig:case_study_pos}
\end{figure*}

\subsection{In-depth Analysis}
\subsubsection{Analysis of Intent Types}\label{subsec:intent_type}
We classify the queries according to their underlying search intent into two categories:
(1) \textbf{Informational Intent}: Natural language questions seeking detailed information or solutions, typically longer and contextual. Queries from the SemEval dataset fall under this category.
(2) \textbf{Exploratory Intent}: Term-based queries aimed at broad exploration with minimal context or structure. Queries from the TREC datasets are categorized here.
Our automatic and human evaluation by intent type indicates that QUIDS is indeed more successful for exploratory tasks than for informational intent tasks. Results are shown in Table~\ref{tab:rouge_intent_type} and Figure~\ref{fig:plot_intent_type}. 
See Appendix~\ref{app:evalaution_intent_type} for detailed analysis.


\begin{figure}[t]
    \centering
    \includegraphics[width=\columnwidth]{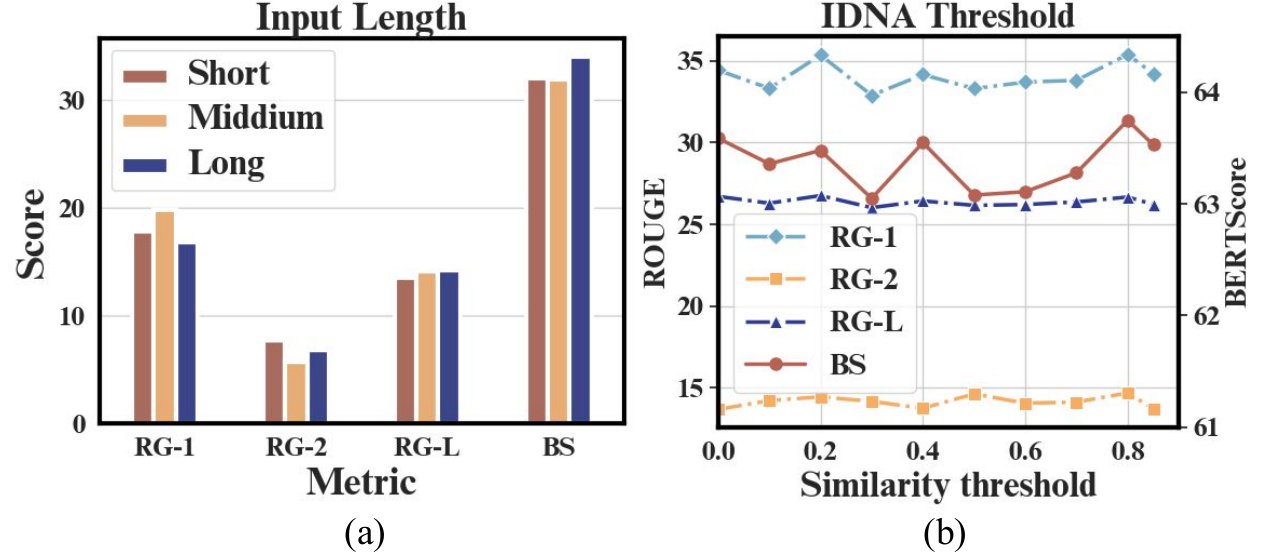}
    \caption{Robustness analysis on (a) input length and (b) IDNA threshold.}
    \label{fig:combi_analysis}
\end{figure}

\subsubsection{Analysis of Document Length}
In real-world scenarios, relevant documents associated with a query can vary in length. Documents over 1024 tokens are truncated in our architecture. 
To assess robustness to input length, we group test documents into short (<512 tokens), medium (512–1024), and long (>1024). 
In the test dataset, 204 are short, 25 medium, and 29 long.
Figure~\ref{fig:combi_analysis}(a) shows that the overall differences in performance among all the input length categories are minimal across all metrics, suggesting that 1024-token inputs are sufficient for inferring the underlying query intent. This may explain why long-document QFS models offer limited gains in this task.

\subsubsection{Sensitivity Analysis of IDNA threshold}\label{subsub:IDNA}
Intent-driven negative augmentation (IDNA) selects irrelevant documents with high similarity to relevant ones as hard negatives for contrastive learning. 
We noticed that most document similarity scores fall between 0.6 and 0.9, making the method robust even with lower thresholds (e.g., <0.6). Performance improves as the threshold increases within this range, with 0.8 yielding the best results, which we use in our experiments.


\subsection{Case Study}
Figure \ref{fig:case_study_pos} illustrates how the model uses cross-attention in the decoder stage to identify irrelevant semantics from a low-ranked document. 
When generating `impacted' without an irrelevant document (Figure \ref{fig:case_study_pos}(a)), the model focuses on economic effects on cars, indicated by `price' in Intent 1.
With an irrelevant document in Figure \ref{fig:case_study_pos}(b), while similar economic attentions 
are observed across both relevant and irrelevant documents when generating the word `effects’ in Intent 2, the model successfully identifies tokens related to prices and cars in relevant documents as irrelevant.
This demonstrates the model's ability to filter out irrelevant content using contrastive learning in the decoder. We include another failure case study in Appendix~\ref{app:case_study}.
\section{Conclusions}

We introduced a novel dual-space modeling approach for the query intent generation task. 
Our approach implements contrastive learning in both encoding and decoding phases, combined with intent-driven hard negative augmentation during data preprocessing, to automatically generate detailed and precise intent descriptions, surfacing what the system likely inferred the query to mean. 
Experimental results show that our model can effectively filter out irrelevant information from the relevant intent space, leading to more accurate intent descriptions than all baselines, including models for Query-Focused Summarization. 
In future work, we plan to improve contextual understanding in distinguishing relevant from irrelevant information and extend our approach to conversational search by mining exploratory needs and explaining the understanding of query intents. Our long-term aim is to improve the transparency in the retrieval process, in particular for exploratory search needs.

\section*{Acknowledgments}
This publication is part of the project LESSEN with project number NWA.1389.20.183 of the research program NWA ORC 2020/21 which is (partly) financed by the Dutch Research Council (NWO).




\section*{Limitations}


\noindent\textbf{Training Efficiency Trade-offs.} Augmenting irrelevant documents enhances robustness but linearly increases training time. We therefore limited negative documents to three per query, which partially alleviates this trade-off but remains suboptimal. To fundamentally resolve this efficiency bottleneck, two promising directions are: (1) adaptive dynamic sampling that prioritizes high-impact negatives through real-time gradient analysis, and (2) curriculum-based augmentation progressively introducing harder negatives as training stabilizes.\\
\noindent\textbf{Dataset Imbalance.} Informational queries dominate the training data over exploratory ones. While our model shows promising performance in exploratory search scenarios, this bias limits deeper intent analysis. Future work should expand out experiments to more datasets, focusing on exploratory queries. One option would be to use LLM-generated synthetic data, specifically creating pseudo-documents that mimic multi-faceted exploratory intents. 
This approach maintains intent modeling consistency while enabling systematic investigation of query complexity, without requiring manual annotation efforts.\\
\bibliography{references}

\appendix

\section{Detailed Comparison with Related Work}\label{app:related_work}
A detailed comparison between our task—Query intent description generation —and related tasks such as Query Understanding (QU), Query-Focused Summarization (QFS), and Pseudo-Relevance Feedback (PRF) is provided in  Table~\ref{tab:comparison_full}.

\section{Experimental Setting Details}\label{app:exp_set}

\subsection{Dataset Details}\label{appendix:dataset-details}

In constructing the Q2ID dataset, documents with multi-graded relevance labels were converted into binary labels, indicating whether each document is relevant to the query.
The dataset is composed of:
\begin{itemize}
  \item 510 entries from TREC tracks (Dynamic Domain 2015–2017, Robust 2004)
  \item 4,878 entries from SemEval-2015/2016 Task 3 on Community Question Answering
\end{itemize}

Each data point is formatted as a quadruple: \textit{<query, relevant documents, irrelevant documents, intent description>}.

The average query length is 7.2 tokens, and the average intent description length is 45.5 tokens.
We follow the original split of Q2ID: 5,000 queries for training, 100 for validation, and 258 for testing.

\subsection{Sensitivity Analysis of Loss Weights}
\label{appendix:loss-weights}

We conduct a sensitivity analysis through a grid search over different combinations of the loss weights ($\lambda_0$, $\lambda_1$, $\lambda_2$). The results of several representative configurations are summarized in Table~\ref{tab:lambda_results}.
\begin{table}[ht]
\centering

\resizebox{\linewidth}{!}
    {
\begin{tabular}{ccc|cccc}
\toprule
$\lambda_0$ & $\lambda_1$ & $\lambda_2$ & RG-1 & RG-2 & RG-3 & BS \\
\hline
-    & -    & -    & \textbf{34.53} & 14.42 & 26.85 & 63.33 \\
0.33 & 0.33 & 0.33 & 33.36 & 14.64 & 26.22 & 63.45 \\
0.20 & 0.20 & 0.60 & 34.51 & \textbf{15.36} & \textbf{27.13} & \textbf{63.52} \\
0.30 & 0.30 & 0.40 & 34.17 & 14.31 & 26.30 & 63.41 \\
0.10 & 0.10 & 0.80 & 32.37 & 14.59 & 25.81 & 62.87 \\
0.20 & 0.40 & 0.40 & 33.20 & 14.56 & 26.10 & 63.22 \\
\bottomrule
\end{tabular}
}
\caption{Sensitivity analysis of loss weights. `-' indicates that the $\lambda$ parameters were treated as learnable during training.}
\label{tab:lambda_results}
\end{table}
We observed that configurations assigning relatively higher weight to $\lambda_2$ tend to yield better overall performance.  
Overly large $\lambda_2$ values (e.g., 0.8) degrade performance by reducing the contribution of the other loss components. 
Configurations that assign more weight to $\lambda_0$ or $\lambda_1$ alone lead to inferior performance, sometimes resulting in no valid generation, and are therefore omitted for clarity. 
Based on these observations, we adopt $(\lambda_0, \lambda_1, \lambda_2) = (0.2, 0.2, 0.6)$ as the default setting.

\begin{table*}[t]
\centering
\resizebox{\linewidth}{!}{
\begin{tabular}{|p{3.2cm}|p{3.6cm}|p{3.6cm}|p{3.6cm}|p{3.6cm}|}
\hline
\textbf{Task} & \textbf{Query Understanding (Classification / Clustering / Expansion)} & \textbf{Query-Focused Summarization (QFS)} & \textbf{Pseudo-Relevance Feedback (PRF)} & \textbf{Our Task: Query Intent Generation (QIG)} \\
\hline
\textbf{Goal} & Predict query intent classes, discover latent topics, or expand queries for better retrieval performance. & Summarize relevant documents to help users consume content. & Refine or reformulate queries to improve retrieval performance. & Generate a natural language description of the search system's inferred intent behind a query.
 \\
\hline
\textbf{Output Form} & Labels (e.g., informational/navigational), clusters, or expanded query terms. & Natural language summary (abstractive or extractive). & Modified query or re-weighted terms. & Natural language explanation of inferred query intent. \\
\hline
\textbf{Use of Irrelevant Documents} & Not used. Focus is on query-only or top-ranked documents. & Rarely used; mainly uses pseudo-relevant documents. & Not used; PRF assumes top-ranked documents are relevant. & Explicitly contrasts relevant and irrelevant documents for intent disentanglement. \\
\hline
\textbf{Application Stage} & Pre-retrieval; typically before document scoring. & Post-retrieval summarization. & Interleaved or pre-retrieval (used for re-ranking or expansion). & Post-retrieval; supporting user query refinement and retrieval debugging \\
\hline
\textbf{User Utility} & Improves ranking accuracy and personalization; not visible to users. & Helps users consume content more efficiently. & Improves recall or relevance through backend query rewriting. & Helps users understand potential mismatches between their intended query meaning and the system’s inferred intent.\\
\hline
\end{tabular}}
\caption{Comparison between Query Understanding (QU) tasks, Query-Focused Summarization (QFS), Pseudo-Relevance Feedback (PRF), and our Query Intent Generation (QIG) task.}
\label{tab:comparison_full}
\end{table*}
\subsection{Implementation}
\label{appendix:impl-details}
To balance efficiency and effectiveness during IDNA augmentation, we set the expected number of irrelevant documents per query to three for training. This results in augmenting 1,984 queries with at least three irrelevant documents each.
Training is conducted for 10 epochs using the Adam optimizer with a learning rate of 0.0001. During decoding, we set a maximum sequence length of 256 tokens and apply beam search with a beam size of 4. We also set a no-repeat n-gram size of 3 to reduce redundancy.

\subsection{Baselines}\label{app:baseline_details}
\textit{Pretrained sequence-to-sequence model baselines}: 
    \textbf{T5} \citep{raffel2020t5}: a Transformer-based encoder-decoder \citep{vaswani2017attention} model trained on a diverse and extensive dataset. We use a pretrained T5-large model that we finetuned on the original Q2ID training dataset.   
    \textbf{BART} \citep{lewis-etal-2020-bart}: also a transformer-based encoder-decoder model, trained by corrupting documents and then optimizing a reconstruction loss. The BART model serves as the backbone of our QUIDS model. 
    
\textit{Query-to-intent description (Q2ID) baseline}: 
    \textbf{CtrsGen} \citep{zhang2020q2id}: a Q2ID model using a bi-directional GRU as encoder architecture. During decoding, it computes contrast scores by considering irrelevant documents to adjust sentence-level attention weights in the relevant documents. 
    
\textit{Large Language Model (LLM) baseline}:
    \textbf{LLaMa3.1-8B-Instruct} \citep{llama31}: 
    We evaluate the LLaMa3.1-8B instruction-tuned text-only model under both zero-shot and two-shot settings and conduct five experimental runs for each setting. For two-shot setting, we randomly using two different examples per run -- one sourced from TREC and the other from SemEval.
    \textbf{OpenAI o3} \citep{openai2025o3}: We evaluate the reasoning-focused OpenAI o3 model. While the model internally generates reasoning tokens, we focus our analysis on the visible output tokens. The same experimental settings and number of runs are used for comparison.
    
\textit{Query Focused Summarization (QFS) baselines}: 
    \textbf{RelReg} \citep{vig2022exploring} and \textbf{RelRegTT} \citep{vig2022exploring}: two-step approaches for QFS consisting of an score-and-rank extractor and an abstractor.
    The extractor is trained to predict ROUGE relevance scores and then the ranked results based on ROUGE are passed to the abstractor. 
    \textbf{SegEnc} \citep{vig2022exploring}: an end-to-end approach tailored for handling longer input texts. SegEnc splits a long input into fixed-length overlapping segments and encodes them separately. The encoding sequences are concatenated so that the decoder can attend to all encoded segments jointly.   
    \textbf{Socratic} \citep{pagnoni-etal-2023-socratic}: an unsupervised, question-driven pretraining approach designed to tailor generic language models for controllable summarization tasks.    
    \textbf{Qontsum} \citep{sotudeh2023qontsum}: an abstractive summarizer that applied Generative Information Retrieval (GIR) techniques. 
    It builds on SegEnc by adding a segment scorer and contrastive learning modules. 
We train the QFS baselines on the Q2ID dataset using the original code provided by the authors, except for Qontsum, which we independently reproduced.

\subsection{Implementation Details for Baselines}
\label{app:implementation_baselines}
RelReg and RelRegTT share the same abstractor, a BART-large model, which also serves as the backbone model for SegEnc, Socratic, and Qontsum.
For RelReg and RelRegTT, we use an input segment length of 1024, whereas SegEnc-based models utilize an input segment length of 512, with a total input length of 4096.
For Socratic training, we use the checkpoint pretrained on Books3 \citep{csaky2020gutenberg} from the Huggingface Model Hub\footnote{https://huggingface.co/Salesforce/socratic-books-30M} and fine-tune it on Q2ID dataset using SegEnc mechanism.
We reproduce the work of Qontsum with the segment length of 512 tokens, temperature of 0.6 and ($\lambda_0=0.6, \lambda_1=0.2, \lambda_2=0.2$) in joint learning.
For all models that divide input text into segments, we apply a 50\% overlap between each segment and its adjacent one.

\begin{table*}[ht]

\centering
\resizebox{\linewidth}{!}
{
\begin{tabular}{l|l|cc|cc|cc|cc|cc}
\toprule[1.5pt]
\multirow{2}{11em}{\textbf{Correlation}} & \multirow{2}{4em}{\textbf{Model}} & \multicolumn{2}{c|}{\textbf{Fluency}} & \multicolumn{2}{c|}{\textbf{Alignment}} & \multicolumn{2}{c|}{\textbf{Inclusion}} & \multicolumn{2}{c}{\textbf{Exclusion}}& \multicolumn{2}{c}{\textbf{Average}} \\
& & $\rho$& $\tau$& $\rho$& $\tau$& $\rho$& $\tau$& $\rho$& $\tau$ & $\rho$ & $\tau$\\
\midrule[1.5pt]


\multirow{2}{11em}{Corr (Human, GPT-4o)} & SegEnc & 0.320& 0.252 & 0.387 & 0.295 & 0.158 & 0.121 & \textbf{0.375}& \textbf{0.309} & 0.310 & 0.244\\
& QUIDS & \textbf{0.476} & \textbf{0.375} & 0.495 &  0.386 & 0.265 & 0.199 & 0.361 & 0.286 & \textbf{0.399} & \textbf{0.312}\\
\midrule
\multirow{2}{11em}{Corr (Human, LLaMa3.1)} & SegEnc & 0.224 & 0.173 & 0.434& 0.329 &  0.153 & 0.116 & 0.343 & 0.272 & 0.289 & 0.223\\
& QUIDS & 0.373& 0.294 & \textbf{0.557}& \textbf{0.424} & \textbf{0.365}& \textbf{0.262}& 0.158& 0.126 & \textbf{0.363} & \textbf{0.276}\\
\bottomrule[1.5pt]
\end{tabular}
}
\caption{Spearman ($\rho$) and Kendall-Tau ($\tau$) correlations between Human evaluation and LLM evaluation of different metrics.}
\label{tab:correlation}
\end{table*}
\begin{table}[!t]
\centering
\resizebox{\linewidth}{!}
{
\begin{tabular}{l|l|cccc}
\toprule[1.5pt]
\textbf{Method} & \textbf{Intent} & \textbf{Fluen.} & \textbf{Align.} & \textbf{Inclu.} & \textbf{Exclu.} \\
\midrule[1.5pt]
\multirow{2}{4em}{Human} & Info. & 4.78 & 3.71& 4.04& 4.80\\
 & Expl. & \textbf{4.89} & \textbf{4.19}& \textbf{4.14} & \textbf{4.81}\\
\midrule
\multirow{2}{4em}{GPT-4o} & Info. & 3.64 & 2.70& 2.63& 3.95\\
& Expl. & \textbf{4.02} & \textbf{3.17}& \textbf{2.96} & \textbf{4.20}\\
\midrule
\multirow{2}{4em}{LLaMa3.1} & Info. & 3.33 & 2.96 & 3.35& 3.99\\
& Expl. & \textbf{3.72} & \textbf{4.11} & \textbf{3.72} & \textbf{4.64}\\

\bottomrule[1.5pt]
\end{tabular}
}
\caption{Comparison of Human and LLM evaluation on informational and exploratory intent types on our model.}
\label{tab:evaluation_intent_type}
\end{table}
\section{Correlation with Human Evaluation}\label{app:correlation}
We assess the correlation between human and LLM evaluators across four qualitative evaluation criteria, presenting the Spearman ($\rho$) and Kendall-Tau ($\tau$) correlations for the best SOTA model SegEnc and our QUIDS model in Table~\ref{tab:correlation}. Overall, our model demonstrates significantly higher human correspondence across all metrics compared to SegEnc, with the exception of the Exclusion score. Correlation performance varies by metric; for Fluency and Factual Alignment—criteria requiring less contextual information—there is a relatively higher degree of agreement with human evaluations. In contrast, the Inclusion and Exclusion scores, which depend on diverse and contextual sources, show lower correlation, suggesting that humans and LLM evaluators adopt different evaluation strategies for more complex criteria. Additionally, we observe that different LLM evaluators exhibit human-like evaluation behaviors across various metrics: LLaMa3.1 shows greater human correspondence in Factual Alignment and Inclusion scores, whereas GPT-4o aligns more closely with human evaluations in Fluency and Exclusion scores.

\section{Evaluation on Intent Types} \label{app:evalaution_intent_type}
In Table~\ref{tab:evaluation}, we observe a human preference over the SegEnc model on metric Factual Alignment, which measures how well the generated query intent description is factually aligned with the ground truth intent. We guess it is due to the model performance difference on different sub-datasets, or on different intent types. And hence we further analyse the evaluations on different intent types. 

\paragraph{Automatic Evaluation}
In the 258 test samples, there are 20 queries with exploratory intents and 238 with informational intents. As shown in Table~\ref{tab:rouge_intent_type}, queries with exploratory intents substantially outperform those with informational intents, achieving 60\% higher ROUGE-2 scores and 20\% higher BERTScores. This indicates that our model is better suited for exploratory queries.
This finding contrasts with the results of \citep{zhang2020q2id}, where the CtrsGen model performed slightly better on the informational SemEval queries than on the exploratory TREC queries. A potential explanation is that the backbone language model used in our approach more freely generates text than the GRU model used in \citep{zhang2020q2id}, particularly when reconstructing complex scenarios for informational intents. This is an aspect that makes our approach more suitable for exploratory search rather than informative search.
\paragraph{Human and LLM Evaluation}
Table~\ref{tab:evaluation_intent_type} presents human and LLM evaluations on our model regarding two intent types. 
In general, exploratory intents consistently outperform informational intents across all metrics. This finding, derived from 50 test samples, aligns with automatic evaluation results on the full test dataset (Table~\ref{tab:rouge_intent_type}).
Figure~\ref{fig:plot_intent_type} shows the evaluation score distribution by intent type on our model, compared to the overall model performance in Figure~\ref{fig:plot_evaluation_all}. The informational intent distribution closely mirrors the overall performance, suggesting that informational queries dominate the dataset and largely influence performance. However, exploratory queries, despite being less frequent, demonstrate superior performance in this task.
When diving into the factual alignment in Figure~\ref{tab:evaluation}, while humans prefer our model for exploratory intent with 4.19 (QUIDS) vs. 3.94 (SegEnc), SegEnc is favored for informative intent with 3.71 (QUIDS) vs. 3.89 (SegEnc). Since informative intent queries dominate, this leads to a lower average score for our model on this metric. These findings indicate that our model is well-suited for exploratory search.

\section{Failure Case Study}\label{app:case_study}
\begin{figure*}[!t]
    \centering
    \includegraphics[width=\textwidth]{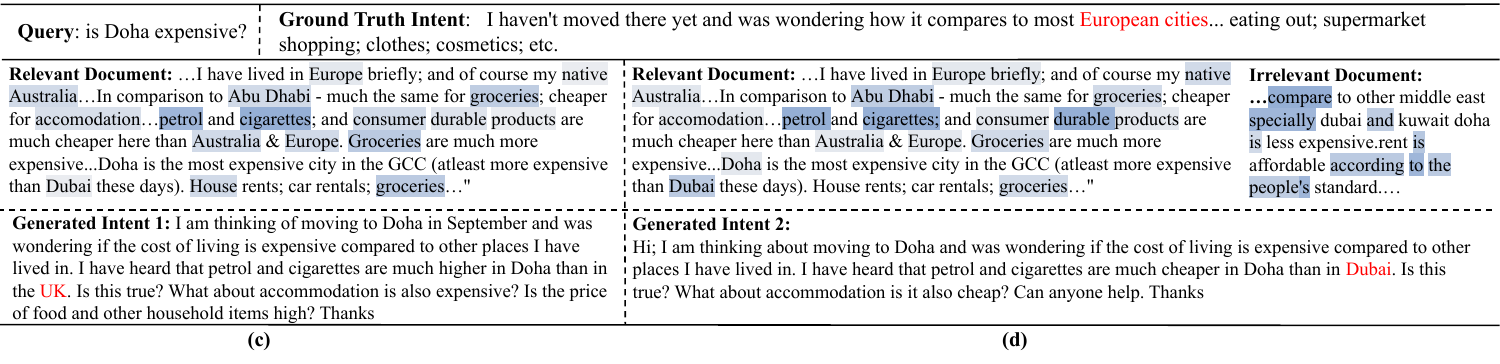}
    \caption{Failure case study. Token-level decoder cross-attention weights for a generated intent token (red) are shown with (c) and without (d) an irrelevant document. Deeper color indicates a higher value.
    }
    \label{fig:case_study_neg}
\end{figure*}
Figure~\ref{fig:case_study_neg} illustrates a failure example of filtering irrelevant information when an irrelevant document is provided. 
The token-level decoder cross-attention weights are compared when generating a content word in the intent, with (c) and without (d) an irrelevant document.
When generating the keyword `UK' and `Dubai', the model mainly focuses on `petrol' and `cigarettes' in the relevant documents for both (c) and (d), which are also contextually important in the generated intent. 
However, the model fails to recognize the relationship between `middle east' in the irrelevance document and `Dubai', leading to the unwanted inclusion of `Dubai' in the intent 2. 
This highlights that our model 
may struggle with excluding information that requires commonsense reasoning or domain-specific knowledge. A direction for future work is to develop advanced approaches that enhance contextual understanding for complex scenarios.
\section{LLM-based Evaluation Details}
\label{appendix_prompt}

Following the method of \citep{liu-etal-2023-g}, we use LLaMa3.1-8B-Instruct\footnote{\url{https://huggingface.co/meta-llama/Llama-3.1-8B-Instruct}.} and GPT-4o\footnote{\url{https://openai.com/index/hello-gpt-4o/}.} as instruction-tuned evaluators to assess the generated intent across four qualitative metrics.
Specifically, we define the evaluation task and criteria, prompting the LLM to generate chain-of-thoughts (CoT) for the `Evaluation Steps'. For LLaMa3.1-8B-Instruct, we use the output token probabilities from the LLMs to normalize the scores and take their weighted summation as the final results:
\begin{equation}
    score = \sum_{i=1}^{n} p(s_i) \times s_i
\end{equation}
where $S=\{s_1, s_2, ..., s_n\}$ represents the predefined score set from the prompt, with a maximum value of 5 in our case. For the close-sourced GPT-4o, we sample 20 times to estimate the token probabilities.
An example prompt for each model is presented below.

\subsection{General Evaluation Prompt}\label{general_prompt}

\textit{You will be given a query, relevant and irrelevant documents with respect to the query. You will also be given a generated query intent description based on the query and documents. The ground truth query intent description will also be provided.} \\
\textit{Your task is to rate the query intent description on one metric.} \\
\textit{Please make sure you read and understand these instructions carefully. Please keep this document open while reviewing, and refer to it as needed.} 
\par
\medskip 
\noindent \textbf{\textit{Evaluation Criteria:}} \\
<<MetricCriteria>>
\par
\medskip 
\noindent \textbf{\textit{Evaluation steps:}} \\
<<EvaluationSteps>> \\

\noindent \textit{Query:}
\begin{verbatim}
{{Query}}
\end{verbatim}
\vspace{6pt}
\noindent \textit{Relevant documents:}

\begin{verbatim}
{{Relevant documents}}
\end{verbatim}
\vspace{6pt}
\noindent \textit{Irrelevant documents:}

\begin{verbatim}
{{Irrelevant documents}}
\end{verbatim}
\vspace{6pt}
\noindent \textit{Generated Intent:}

\begin{verbatim}
{{Generated intent}}
\end{verbatim}
\vspace{6pt}
\noindent \textit{Ground Truth Intent:}

\begin{verbatim}
{{Gound truth intent}}
\end{verbatim}
\vspace{6pt}

\noindent \textbf{\textit{Evaluation Form (scores ONLY):}}\\
\textit{- <<MetricName>>:}

\subsection{Evaluation Prompt on Fluency}\label{fluent_prompt}

\noindent \textbf{\textit{Evaluation Criteria:}} \\
\textit{Fluency Score (1-5) - This metric measures if the generated query intent description reads naturally, understandably, and without noticeable errors or disruptions.}
\par
\medskip 
\noindent \textbf{\textit{Evaluation steps:}} \\
\textit{1. Carefully review the provided query, relevant, and irrelevant documents to understand the context and content.} \\
\textit{2. Read the ground truth query intent description to understand the ideal response. This serves as a benchmark for evaluating the fluency of the generated description.}\\
\textit{3. Carefully read the generated query intent description. Focus on the fluency aspect, considering factors such as grammatical correctness, naturalness, clarity, coherence, and readability.}\\
\textit{Assign a rating from 1 to 5 based on the level of fluency.} \\

\noindent \textbf{\textit{Evaluation Form (scores ONLY):}}\\
\textit{- Fluency:}

\subsection{Evaluation Prompt on Factual Alignment}\label{alignment_prompt}

\noindent \textbf{\textit{Evaluation Criteria:}} \\
\textit{Factual Alignment (1-5) - This metric measures if the generated query intent description is factually aligned with the ground truth intent. Ensuring the facts presented in the generated description are correct and match those in the ground truth description. Verifying that all key facts and points mentioned in the ground truth are covered in the generated description without omission. Any hallucination that diverges from the ground truth should be flagged.}
\par
\medskip 
\noindent \textbf{\textit{Evaluation steps:}} \\
\textit{1. Review the ground truth intent description for the central facts and points that convey the query's purpose.} \\
\textit{2. Read the generated intent description and list the main facts and points it conveys.}\\
\textit{3. Compare the lists from the ground truth and generated descriptions for consistency in content. Look for alignment in terms of content, completeness, and accuracy.}\\
\textit{4. Identify any key facts or points from the ground truth that are missing in the generated description (omissions) and note any information in the generated description that is not present or diverges from the ground truth (hallucinations).}\\
\textit{Assign a rating from 1 to 5 based on the level of factual alignment.} \\

\noindent \textbf{\textit{Evaluation Form (scores ONLY):}}\\
\textit{- Factual Alignment:}

\subsection{Evaluation Prompt on Inclusion Score}\label{inclusion_prompt}
\noindent \textbf{\textit{Evaluation Criteria:}} \\
\textit{Inclusion Score (1-5) - This metric measures how well the generated query intent includes important details from the query and relevant documents. Assessing whether the generated description captures key elements that are directly relevant to the query. Evaluating if the generated description thoroughly includes significant points from the relevant documents. Ensuring that the included details are integrated in a way that maintains the context and importance as presented in the relevant documents.}
\par
\medskip 
\noindent \textbf{\textit{Evaluation steps:}} \\
\textit{1. Review the query and relevant documents to extract the main facts, significant points, and key elements that directly address the query.} \\
\textit{2. Read the generated query intent description and list the key details it includes.}\\
\textit{3. Compare the key details and elements from the generated description with those identified from the query and relevant documents, checking for inclusion and alignment.}\\
\textit{4. Assess how well the included details are integrated into the generated description, ensuring they maintain the context and importance as presented in the relevant documents.}\\
\textit{Assign a rating from 1 to 5 based on the thoroughness and relevance of the included details.} \\

\noindent \textbf{\textit{Evaluation Form (scores ONLY):}}\\
\textit{- Inclusion Score:}

\subsection{Evaluation Prompt on Exclusion Score}\label{exclusion_prompt}
\noindent \textbf{\textit{Evaluation Criteria:}} \\
\textit{Exclusion Score (1-5) - This metric measures if the generated query intent description excludes information present in the irrelevant documents that is not relevant to the query and relevant documents. Evaluating whether the description effectively filters out information that is irrelevant to the query. Ensuring that the description does not include misleading or incorrect information found in the irrelevant documents. Evaluating whether the description effectively filters out information present in the irrelevant documents but focus on topics different from those in relevant documents.}
\par
\medskip 
\noindent \textbf{\textit{Evaluation steps:}} \\
\textit{1. Carefully read through the irrelevant documents to pinpoint details, facts, or topics that are not relevant to the query and relevant documents.} \\
\textit{2. Read the generated query intent description and extract the key details and points included in the description.}\\
\textit{3. Compare the extracted content from the generated description with the irrelevant information identified in the irrelevant documents to check for the presence of any irrelevant details.}\\
\textit{4. Assess how effectively the generated description filters out irrelevant information, ensuring it focuses only on the query and relevant documents.}\\
\textit{Assign a rating from 1 to 5 based on the level of exclusion of irrelevant details.} \\

\noindent \textbf{\textit{Evaluation Form (scores ONLY):}}\\
\textit{- Exclusion Score:}

\end{document}